\documentclass[
aps,
jcp,
showpacs
amsmath,
amssymb,
floatfix,
letterpaper,
lengthcheck,
superscriptaddress
]{revtex4-1}

\usepackage[version=3]{mhchem} 
 
\usepackage{rotating}
\usepackage{amsmath}
\usepackage{bbm}
\usepackage{subfigure}
\usepackage{color}
\usepackage{braket}
\usepackage{dsfont}

\newcommand*{\abs}[1]{\mathopen| #1 \mathclose|}

\newcommand{\be}[1]{\begin{equation}#1\end{equation}}

\newcommand{\ba}[1]{\begin{align}#1\end{align}}

 \newcommand{\brakete}[2]{\left< #1 \right|\left. #2 \right>} 
 
\newcommand{\pd}[2]{\frac{\partial #1}{\partial #2}} 
 
\newcommand{\tb}[1]{\textbf{#1}} 

\begin{document}

\title{Non-adiabatic and time-resolved photoelectron spectroscopy for molecular systems}

\author{Johannes Flick}
  \email[Electronic address:\;]{flick@fhi-berlin.mpg.de}

\author{Heiko Appel}
  \affiliation{Fritz-Haber-Institut der Max-Planck-Gesellschaft, Faradayweg 4-6, D-14195 Berlin-Dahlem, Germany}
\author{Angel Rubio}
  \affiliation{Fritz-Haber-Institut der Max-Planck-Gesellschaft, Faradayweg 4-6, D-14195 Berlin-Dahlem, Germany}
  \affiliation{Nano-Bio Spectroscopy group, Dpto. F\'isica de Materiales, Universidad del Pa\'is Vasco, Centro de F\'isica de Materiales CSIC-UPV/EHU-MPC and DIPC, Av. Tolosa 72, E-20018 San Sebasti\'an, Spain}

\date{\today}

\begin{abstract}
We quantify the non-adiabatic contributions to
the vibronic sidebands of equilibrium and explicitly time-resolved
non-equilibrium photoelectron spectra for a vibronic model 
system of Trans-Polyacetylene. Using exact diagonalization, we directly evaluate 
the sum-over-states expressions for the linear-response 
photocurrent. We show that spurious peaks appear in the Born-Oppenheimer
approximation for the vibronic spectral function, which are 
not present in the exact spectral function of the system. The effect can be traced back to the 
factorized nature of the Born-Oppenheimer initial and final photoemission
states and also persists when either only initial, or final states are replaced 
by correlated vibronic states. Only when correlated initial and final 
vibronic states are taken into account, the spurious spectral weights of the Born-Oppenheimer approximation
are suppressed. In the non-equilibrium case, we illustrate for an initial 
Franck-Condon excitation and an explicit pump-pulse excitation how the 
vibronic wavepacket motion of the system can be traced in the time-resolved 
photoelectron spectra as function of the pump-probe delay. 
\end{abstract}


\date{\today}

\maketitle


\section{Introduction}
\label{sec:Intro}

Photoelectron spectroscopy is a well-established experimental method 
to probe the structure of atoms, molecules and solids \cite{Huefner2003,Reinert2005}. In comparison to other 
spectroscopic methods such as optical-absorption spectroscopy, photoelectron 
spectroscopy is based on non-neutral transitions between many-body states: initial
and final states, which have vanishing matrix elements for charge-neutral transitions, 
might have non-vanishing matrix elements for non-neutral transitions.
Hence, photoelectron spectroscopy allows to observe dipole or quadrupole forbidden 
transitions, which would otherwise not be accessible in optical-absorption 
spectroscopy.\\ 
With the appearance of femtosecond laser pulses \cite{Pietzsch2008}, which lead 
to the Nobel prize in chemistry awarded to A. H. Zewail, the field 
of time-resolved photoelectron spectroscopy has seen tremendous developments in 
recent years. Femtosecond laser pulses are routinely used to 
probe a large variety of intra-molecular effects. By combining short pulses with 
the new possibilities of time-resolved photoelectron spectroscopy, experimentalists 
are now able to realize femtosecond pump-probe photoelectron spectroscopy: Here, two 
independent laser pulses are employed to eject photoelectrons. The first pulse is 
used to excite the sample, followed by a second laser pulse after a finite delay 
time. The second laser pulse photoexcites the system to emit a photoelectron. The energy 
and angle resolved distribution of photoelectrons can then be detected by 
the measurement apparatus. Tuning the delay time allows to monitor 
dynamical processes in the system.\\
These novel techniques for time-resolved pump-probe photoelectron spectroscopy have 
already been used to experimentally study and characterize ultrafast photochemical 
dynamic processes in liquid jets \cite{Buchner2010}, to follow ultrafast electronic 
relaxation, hydrogen-bond-formation and dissociation dynamics \cite{Yamaguchi2000}, 
to probe unimolecular and bimolecular reactions in real-time \cite{Motzkus1996}, or 
to investigate multidimensional time-resolved dynamics near conical 
intersections \cite{Hauer2007}, all on a femtosecond timescale, to mention a few.\\
Driven by such novel experimental possibilities, there is an ongoing demand 
to extend and refine existing theory to allow for a first-principles description of 
time-dependent pump-probe photoelectron experiments and to treat the electronic and ionic responses on an equal footing. 
Along these lines, first steps 
have already been taken, which focus on photoelectron spectroscopy in real-time. On 
the level of time-dependent density functional theory (TDDFT), the first pioneering work to
describe photoelectron spectra was based on the momentum distribution of the Kohn-Sham orbitals
recorded at a reference point far from the system \cite{Pohl2000}. More recently a mask 
technique has been developed \cite{DeGiovannini2012} and extended to attosecond pump-probe
spectroscopy \cite{DeGiovannini2013}. This approach captures the 
time-dependent case intrinsically  and allows e.g.~to directly 
simulate the photoemission process for given delay times and shapes of pump and probe 
pulses. However, in this approach, the description is so far limited to classical nuclei. 
As a result, the vibrational sidebands in time-resolved photoelectron spectra and photoabsorption are not fully
captured. Other approaches in similar direction have been realized combining TDDFT and ab-initio molecular 
dynamics \cite{Ren2013}, or by using an reduced density matrix description, which also 
relies on the Born-Oppenheimer approximation \cite{Dutoi2013}. \\
On the other hand, approaches based on the Born-Oppenheimer approximation allow 
for a detailed analysis of angular resolved photoelectron spectra and for a reconstruction of molecular orbital densities  directly from the spectra \cite{Puschnig2009,Dauth2011}. Standard quantum chemical approaches for photoelectron spectroscopy as e.g.~the 
double-harmonic approximation (DHA) \cite{Koziol2009} allow to capture vibrational 
sidebands through Franck-Condon factors \cite{Cederbaum1974}. Here, the vibronic 
nature of the involved initial and final states is taken into account within a 
harmonic approximation of the corresponding Born-Oppenheimer surfaces.
Although the vibrational sidebands of photoelectron spectra can be approximately 
captured in the DHA, such an approach lacks the possibility to describe 
time-resolved pump-probe experiments explicitly.\\
In this work, we attempt to compare and validate existing computational tools for photoelectron
spectroscopy of vibronically coupled systems. We present an approach for time-resolved 
photoelectron spectra, which explicitly includes the vibronic nature of the involved 
states and allows to follow the photoemission process in real-time. For a realistic model 
system of small Trans-Polyacetylene oligomer chains, we investigate time-resolved 
photoelectron spectra and compare to approaches such as the double-harmonic approximation.
Our study is based on exact diagonalization of vibronic Hamiltonians and real-time
propagations of the time-dependent Schr\"odinger equation in the combined electronic 
and vibrational Fock space of the system. This procedure gives us access to the exact correlated 
electron-nuclear eigenstates and time-evolved states of the system and enables us to 
test different levels of approximations against our correlated reference calculations. 
In particular, we focus on non-adiabatic effects beyond the Born-Oppenheimer approximation.\\
The paper is organized as follows: In section~\ref{sec:model_system}, we introduce our employed
model for Trans-Polyacetylene oligomers and provide a comparison of the 
Born-Oppenheimer states of the model to the exact correlated energy eigenstates of the Hamiltonian.
Different levels of approximations for the photocurrent are introduced in section \ref{sec:spectral}
and the relation of photoelectron spectra to the one-body spectral function is discussed. In 
addition, we focus on time-resolved pump-probe photoemission spectroscopy.
In section~\ref{sec:results}, we apply the theoretical tools of section \ref{sec:spectral} to
our model system for Trans-Polyacetylene and discuss spurious peaks, which appear in
the Born-Oppenheimer approximation for the spectra. To illustrate our approach for explicitly
time-resolved spectroscopy, we numerically simulate two pump-probe photoelectron experiments: 
as first example, we consider a Frank-Condon transition as excitation mechanism and in a second 
example, we explicitly propagate the system in the presence of a pump pulse. In both cases, 
photoelectron spectra are recorded and we highlight the underlying nuclear wavepacket motion
and the differences to the equilibrium spectra. Finally, in section~\ref{sec:conc} we 
summarize our findings and give an outlook for future work.

\begin{figure}[t]
  \begin{center}
    \includegraphics[width=0.48\textwidth]{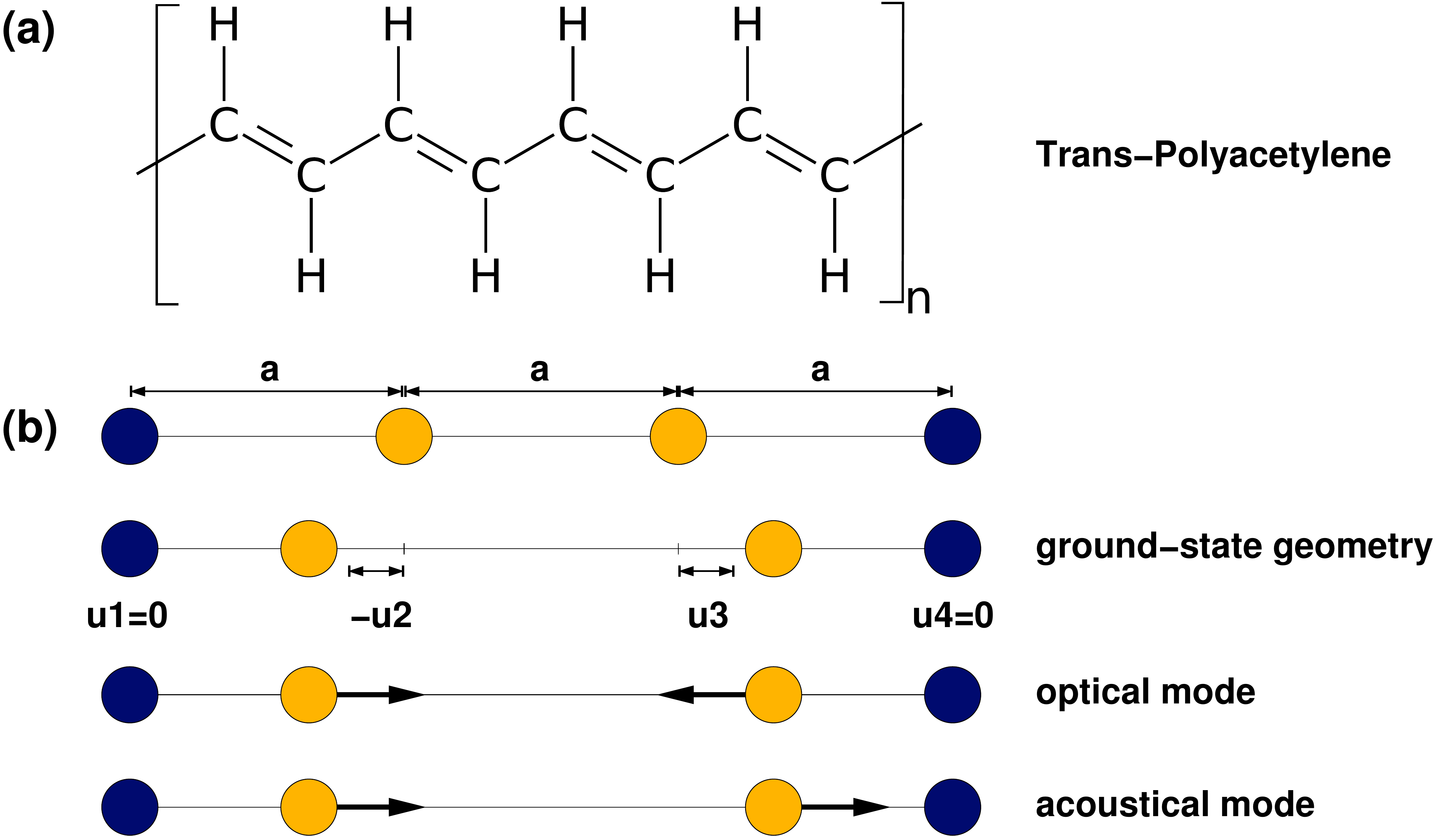}
  \end{center}
  \caption{(a) Chemical structure of Trans-Polyacetylene. 
           (b) Model system for SSH-chain: four Trans-Polyacetylene oligomers with 
               clamped ends. The coordinates $u_j$ describe the shift of the oligomers 
               with respect to a perfectly periodic arrangement of lattice spacing $a$. 
               Both, in the exact Born-Oppenheimer and in the exact correlated ground state, 
               the chain favors a dimerized arrangement. Also shown 
               are the optical and acoustical phonon modes of the chain.
  }
  \label{fig:4-site-chain}
\end{figure}

\section{Model system}
\label{sec:model_system}

\subsection{Su-Schrieffer-Heeger-Hamiltonian and exact eigenvalues and eigenfunctions}
\label{sec:hamiltonian}

In this section, we briefly review the Su-Schrieffer-Heeger (SSH) model \cite{Su1979,Heeger1988} for 
Trans-Polyacetylene (PA) and the exact-diagonalization approach.\\
To model PA oligomer chains (Fig.~\ref{fig:4-site-chain} (a)),
we employ the SSH Hamiltonian to describe $\pi$-electrons in a polymer chain %
\ba{\hat{H}_{ssh}&=\hat{H}_\pi+\hat{H}_{ph}+\hat{H}_{\pi-ph}\label{eq:ssh1}\\
  \hat{H}_\pi&=-T\sum\limits_{n,\sigma}\hat{c}^\dagger_{n+1,\sigma}\hat{c}_{n,\sigma}+\hat{c}^\dagger_{n,\sigma}\hat{c}_{n+1,\sigma} \nonumber\\
  \hat{H}_{ph}&=\sum\limits_{n}\frac{\hat{p}^2_{n}}{2M} +\frac{K}{2} \left(\hat{u}_{n+1}-\hat{u}_{n}\right)^2 \nonumber\\
  \hat{H}_{\pi-ph}&=\sum\limits_{n,\sigma} \alpha \ (\hat{u}_{n+1}-\hat{u}_n)\left(\hat{c}^\dagger_{n+1,\sigma}\hat{c}_{n,\sigma}+\hat{c}^\dagger_{n,\sigma}\hat{c}_{n+1,\sigma}\right)\nonumber
.}
With $\hat{c}^\dagger_{n,\sigma}$, and $\hat{c}_{n,\sigma}$, we denote the usual fermionic 
creation and annihilation operators, which create or destroy $\pi$-electrons with spin 
$\sigma$ on site $n$ of the chain. The nuclear subsystem in the Hamiltonian is described 
by the nuclear displacement operators $\hat{u}_n$ and the nuclear momentum operators 
$\hat{p}_n$. Expectation values of the operator $\hat{u}_n$ measure the displacement 
of the nuclear positions of site $n$ with respect to an equidistant arrangement of 
the oligomers in the chain. The displacement and momentum operators obey the usual 
bosonic commutation relations, 
$[\hat{p}_i, \hat{p}_j] = 0,\, [\hat{u}_i, \hat{u}_j] = 0,\, [\hat{u}_j, \hat{p}_j] = i\hbar$.
For clarification, we always use the hat symbol $\left[ \ \hat{} \ \right]$ 
to distinguish between quantum mechanical operators and classical variables.
Throughout the paper, we use the standard set of parameters for the 
SSH-Hamiltonian \cite{Heeger1988}:
$\alpha$=4.1 eV/\AA, $T$=2.5 eV, $K$=21 eV/\AA$^2$, $M$=1349.14 eVfs$^2$/\AA$^2$, 
which leads to a lattice spacing of $a$=1.22 \AA~in the chain. For this set of parameters, the chain energetically favors a dimerized arrangement of the oligomers in the ground-state, leading to a 
nonvanishing displacement coordinate $u\ne0$. The dimerization is illustrated 
in Fig.~\ref{fig:4-site-chain}. \\
The SSH Hamiltonian has been used in the literature to describe soliton 
propagation in conjugated polymers \cite{Heeger1988}, or to study coupled 
electron-nuclear dynamics \cite{Stella2011,Franco2013,Franco2013a}.
The Hamiltonian in Eq.~\ref{eq:ssh1} can be divided into three parts: (1) the electronic Hamiltonian 
$\hat{H}_{\pi}$, which models electron hopping of $\pi$-electrons within a tight binding scheme. (2) The 
nuclear Hamiltonian $\hat{H}_{ph}$ describes all nuclei as a chain of coupled 
quantum harmonic oscillators, and (3) the interaction part in the Hamiltonian 
$\hat{H}_{\pi-ph}$ takes the coupling of electrons and nuclei up to first-order 
in the nuclear displacement into account. The electron-phonon coupling $\hat{H}_{\pi-ph}$
may be combined with the kinetic term $\hat{H}_{\pi}$ of the $\pi$ electrons. The hopping
parameter $-T$ is then replaced by $-T + \alpha \ (\hat{u}_{n+1}-\hat{u}_n)$. Physically
speaking, it is more likely for electrons to hop when two nuclear positions approach each other
or conversely the effective hopping parameter is decreased when the nuclei are moving apart.\\
To get access to all eigenvalues and eigenstates of the system,
we employ an exact diagonalization technique \cite{Streltsov2010,Kaneko2013}. In the combined electron-nuclear Fock space, we explicitly 
construct matrix representations for all operators present in Eq.~\ref{eq:ssh1}. 
For the photoelectron spectra, we choose to work in Fock space, since here we have directly access to states with different electron number $N$.
The matrix representations for the electronic creation and annihilation operators 
are constructed in terms of a Jordan-Wigner transformation \cite{Jordan1928} and 
the nuclear position and momentum operators are represented on a uniform real-space 
grid applying an 8th-order finite-difference scheme. In the present work, we use a two-dimensional phonon grid with 35x35 grid points. Hence, the total Fock space containing up to eight electrons has the size $M_{\rm tot}=4^4$x35x35 = 313600. The four (three) electron Hilbert space has a size of $M^{[4]([3])}_{\rm max} =$ 70(56)x35x35 = 85750(68600) basis functions.\\
The Hamiltonian in Eq.~\ref{eq:ssh1} commutes with the spin operators $\hat S_z$,
$\hat S^2$, particle number $\hat N$, and parity $\hat P$.
By exploiting all these symmetries, we first block-diagonalize the 
Hamiltonian by ordering basis states according to tuples of eigenvalues of 
all symmetry operators that commute with the Hamiltonian. All remaining blocks 
in the Hamiltonian are then diagonalized with a dense eigenvalue solver. In contrast
to standard sparse diagonalization approaches for exact diagonalization, this 
procedure gives us access to the {\em full spectrum} of {\em all} $M_{\rm tot}$ eigenvalues and eigenvectors 
of the static Schr\"odinger equation%
\begin{align}
\label{eq:FullStaticSE}
\hat{H}_{ssh} \ket{\Psi^{(N)}_j} = E^{(N)}_j \ket{\Psi^{(N)}_j}. 
\end{align} 
Here, the eigenstates $\ket{\Psi^{(N)}_j}$ and eigenvalues $E^N_j$ of the SSH Polyacetylene
chain refer to the exact correlated stationary states of the combined system of electrons 
and nuclei in Fock space.
To simplify the following discussion of electron removal, we always indicate the
number of electrons explicitly with superscript $N$.\\
To solve for the time evolution of arbitrary initial states $\ket{\Phi^{(N)}(0)}$ 
in the presence of pump and probe pulses, we explicitly propagate the 
time-dependent Schr\"odinger equation
\begin{align} \label{eq:td-schroed}
i\hbar \pd{}{t} \ket{\Phi^{(N)}(t)} = \hat{H}_{ssh} \ \ket{\Phi^{(N)}(t)},
\end{align}
with a Lanczos propagation scheme \cite{Park1986,HocL97}.
In the following, the exact diagonalization of the static Schr\"odinger equation
and the time-evolved states of the correlated system serve 
as exact reference to test the quality and validity of approximate schemes
for photoelectron spectra. Due to the exponential scaling of the 
Fock space size, the calculation of exact eigenstates and time-evolved 
wavefunctions is limited to small SSH chains (maximum of four oligomers in the present
case). Although the exact numerical solutions are only available for small SSH chains,
they serve as valuable reference to test approximate schemes, which then can be employed
for larger systems.

\begin{figure}[t]
  \begin{center}
    \includegraphics[width=0.48\textwidth]{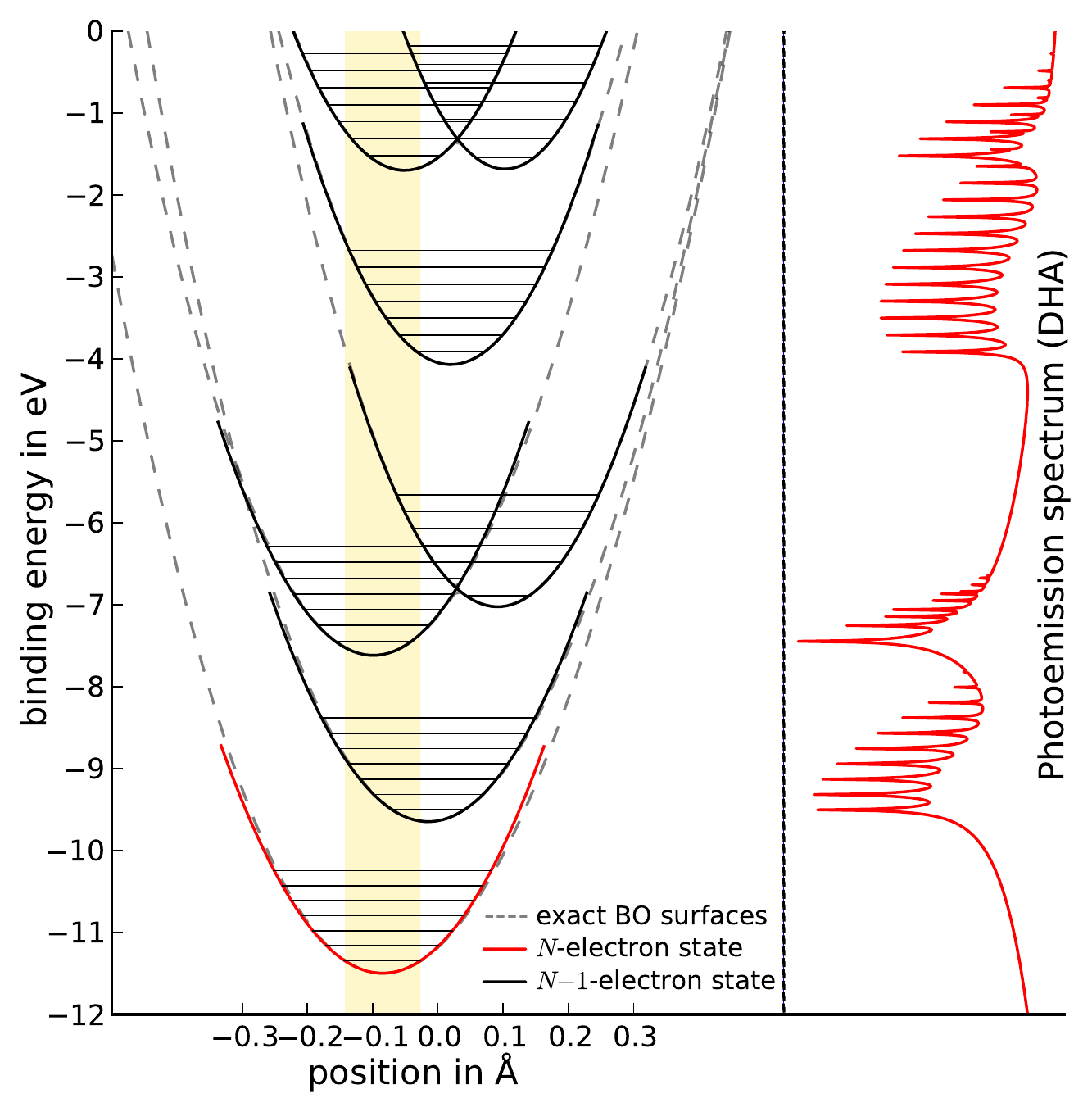}
  \end{center}
  \caption{
    Exact Born-Oppenheimer surfaces and harmonic approximation for 
    a Trans-Polyacetylene oligomer: 
    In the panel on the left-hand side, an one-dimensional cut along 
    the optical axis of the exact potential energy surfaces is shown 
    in dashed lines. The harmonic approximations to the exact BO surfaces 
    are shown in solid lines, where black lines refer to $N-1$ electron 
    states and the red line corresponds to the $N$-electron ground state. 
    In the panel on the right hand side, the corresponding photoelectron 
    spectrum in double-harmonic approximation is shown.}
  \label{fig:cartoon-dha}
\end{figure}
%

\begin{center}
  \begin{table}
  \renewcommand{\tabcolsep}{2.5mm}
    \begin{tabular}{| c| c | c | c |c |}
\hline
state \# & $E^{\text{exact}}$  &  $E^{\text{exact}}_{BO}$ &  (e,o,a)    & overlap\\\hline\hline
1   &    -11.3414      &  -11.3419      &  1,0,0  & 0.9986 \\
2   &    -11.2166      &  -11.2171      &  1,0,1  & 0.9986 \\ 
3   &    -11.1583      &  -11.1588      &  1,1,0  & 0.9955\\ 
4   &    -11.0918      &  -11.0924      &  1,0,2  & 0.9986\\ 
5   &    -11.0336      &  -11.0341      &  1,1,1  & 0.9955\\
86  &    -9.5155       &   -9.5157      &  1,10,0 & 0.9676 \\
87  &    -9.5076       &   -9.5078      &  1,8,3  & 0.9740\\
\hline
\hline
state \# &  $E^{\text{exact}}$  &  $E^{\text{harmonic}}_{BO}$  &  (e,o,a) & overlap\\\hline\hline
1       &-11.3414&        -11.3419      &  1,0,0  & 0.9986 \\
2       &-11.2166&        -11.2171      &  1,0,1  & 0.9986 \\
3       &-11.1583&        -11.1587      &  1,1,0  & 0.9953 \\
4       &-11.0918&        -11.0923      &  1,0,2  & 0.9986 \\
5       &-11.0336&        -11.0339      &  1,1,1  & 0.9953 \\
86      &-9.5155 &         -9.5102      &  1,10,0 & 0.8964 \\
87      &-9.5076 &         -9.5023      &  1,8,3  & 0.9361\\

\hline
    \end{tabular}
  \caption{Exact correlated energies $E^{\text{exact}}$ , BO energies $E^{\text{exact}}_{BO}$ and $E^{\text{harmonic}}_{BO}$ and overlap between exact and 
BO states. All energies are given in eV. The label (e,o,a)  refers to 
the BO quantum number of the state (electronic state, optical phonon mode, acoustical phonon mode).
Note, that the exact BO energies $E^{\text{exact}}_{BO}$ provide a lower bound to the exact 
correlated energies $E^{\text{exact}}$.}
  \label{tab:overlap}
  \end{table}
\end{center}

\subsection{Born-Oppenheimer approximation for the Su-Schrieffer-Heeger-Hamiltonian}
\label{sec:boa}

To introduce the required notation for the following sections and to illustrate the
exact potential energy surfaces, we briefly
discuss the Born-Oppenheimer approximation for the SSH model.
By setting the nuclear kinetic energy in Eq.~\ref{eq:ssh1} to zero, the
nuclear displacements become classical parameters and we arrive at the
electronic Born-Oppenheimer Hamiltonian for the SSH chain
\ba{\hat{H}_{ssh,el}&=\hat{H}_{\pi,el}+\hat{H}_{\pi-ph,el}\label{eq:el}\\
  \hat{H}_{\pi,el}&=-T\sum\limits_{n,\sigma}\hat{c}^\dagger_{n+1,\sigma}\hat{c}_{n,\sigma}+\hat{c}^\dagger_{n,\sigma}\hat{c}_{n+1,\sigma} \nonumber\\
  \hat{H}_{\pi-ph,el}&=\sum\limits_{n,\sigma} \alpha \ (u_{n+1}-u_n)\left(\hat{c}^\dagger_{n+1,\sigma}\hat{c}_{n,\sigma}+\hat{c}^\dagger_{n,\sigma}\hat{c}_{n+1,\sigma}\right)\nonumber
}
with the corresponding eigenvalue problem
\be{\label{eq:boa_eigenvalue_problem}
  \hat{H}_{ssh,el}\left(\{u_n\}\right)\ket{\phi_j^{(N)}\left(\{u_n\}\right)} = \epsilon_{el,j}\left(\{u_n\}\right) \ket{\phi_j^{(N)}\left(\{u_n\}\right)}
.}
The eigenvalues $\epsilon_{el,j}\left(\{u_n\}\right)$ as function of the classical
coordinates $\{u_n\}$ denote the Born-Oppenheimer 
surfaces of the system. Similar to the case of the exact correlated eigenstates $\ket{\Psi^{(N)}_j}$, 
we here use for electronic Born-Oppenheimer states $\ket{\phi_j^{(N)}}$ a superscript $(N)$ 
to distinguish between the Hilbert spaces of different electron numbers. In analogy to the exact diagonalization
approach for the full Hamiltonian, as discussed in the previous section, we employ
here a dense exact diagonalization scheme for the electronic Born-Oppenheimer Hamiltonian. This procedure
gives us access to {\em all} exact Born-Oppenheimer surfaces and corresponding Born-Oppenheimer states 
of the SSH chain. In addition to
the exact surfaces, we compute the Hessian of the electronic energies with respect
to the displacements. By diagonalizing the Hessian, we arrive at the harmonic
approximation for the Born-Oppenheimer surfaces. In Fig.~\ref{fig:cartoon-dha}, we illustrate
the exact potential energy surfaces (dashed lines) along the axis of the optical normal mode 
of the chain and compare to the harmonic approximation of the surfaces (solid lines). As can 
be seen from the figure, for the Hamiltonian in Eq.~\ref{eq:ssh1} the harmonic approximation is very close to the exact surfaces. Overall,
the model behaves rather harmonic and only small anharmonicities are present. We emphasize, that
the almost harmonic nature of the exact potential energy surfaces originates from the quadratic
interaction term in the phonon Hamiltonian $\hat{H}_{ph}$, since already in the exact model 
Hamiltonian only quadratic nuclear interaction terms are included. The only source of anharmonicity
and non-adiabaticity is the electron-phonon coupling term $ \hat{H}_{\pi-ph}$ in Eq.~\ref{eq:ssh1}, 
which introduces only small anharmonicities and non-adiabatic couplings between different 
electronic surfaces.\\
For each fixed set of nuclear 
coordinates $\{u_n\}$, the electronic eigenstates $\ket{\phi_j^{(N)}\left(\{u_n\}\right)}$ 
form a complete set in the many-particle Hilbert space of the electrons. For 
a given set of nuclear displacement coordinates $\{u_n\}$, we can expand the 
exact many-body wavefunction in terms of the electronic eigenstates 
$\ket{\phi_j^{(N)}}$ and the nuclear eigenstates $\ket{\chi_{ij}}$ in terms of
the Born-Huang expansion \cite{Born1956}:
\ba{\label{Born-Huang}\ket{\Psi_i^{(N)}\left(\{u\}\right)}=&\sum\limits_{j=1}^\infty\ket{ \chi_{ij}\left(\{u_n\}\right)  \otimes \phi_j^{(N)}\left(\{u_n\}\right)}\\
=&\sum\limits_{j=1}^\infty\ket{ \chi_{ij}\left(\{u_n\}\right)  \phi_j^{(N)}\left(\{u_n\}\right)}\nonumber
.}
Here, the electronic eigenstates $\ket{\phi_j^{(N)}}$
depend parametrically on the oligomer displacements $\{u_n\}$ and the nuclear
eigenstates $\ket{\chi_{ij}}$ are functions of $\{u_n\}$.
We solve for the states $\ket{\chi_{ij}\left(\{u_n\}\right)}$ 
by diagonalizing the corresponding nuclear Born-Oppenheimer Hamiltonian
\ba{\hat{H}_{ssh,ph,j}&=\hat{H}_{ph,j}+\epsilon_{el,j}\left(\{u_n\}\right) 
}
directly in the real-space representation.
In Tab.~\ref{tab:overlap}, we compare for the lowest five states and two higher-lying states
the exact BO energies $E^{\text{exact}}_{BO}$ and the BO energies in harmonic approximation 
$E^{\text{harmonic}}_{BO}$ to the exact many-body energies of the correlated system $E^{\text{exact}}$.
In addition, we give the overlaps of Born-Oppenheimer and exact states.
For the present model, low lying Born-Oppenheimer states in harmonic
approximation and exact Born-Oppenheimer states are both very good approximations to the exact correlated states. 
In particular, the exact BO ground state has an overlap of 99.86\% with the exact correlated ground state. For 
higher-lying states, the harmonic approximation of the potential energy surfaces yields states with
less accurate energies and overlaps compared to the exact BO states.
Despite the good agreement for the low lying states, we demonstrate in section~\ref{sec:results}
that the differences between exact and harmonic BO and exact correlated states for higher-lying 
levels cause sizeable deviations between the exact and the corresponding exact or harmonic BO 
photoelectron spectra. In particular, the BO photoelectron spectra acquire 
spurious peak amplitudes that are not present in the exact correlated spectrum.

\section{Theory of static and time-dependent photoelectron spectroscopy} 
\label{sec:spectral}

In this section, we briefly review the connection between photoelectron spectra 
and the one-body spectral function known from literature \cite{Hedin1999,Almbladh1983,Almbladh2006,Uimonen2013}
and extend the discussion to vibronic states. For later purposes,
we discuss the equilibrium and nonequilibrium spectral functions.
Since our emphasis in the present 
work is on pump-probe photoelectron experiments for vibronic systems, 
we keep an explicit focus on the reference-state dependence of the vibronic 
one-body spectral function and we discuss how the photocurrent can be expressed 
in a real-time evolution. \\
In terms of Fermi's Golden Rule, we can formulate the exact expression for 
the photocurrent $J_\tb{k}(\omega)$ in first order perturbation theory \cite{Hedin1999,Onida2002,Reinert2005} as
\be{J_\tb{k}(\omega)=\frac{2\pi}{\hbar}\sum\limits_j\abs{\bra{\Psi^{(N)}_{j,\tb{k}}}\hat{\Delta}\ket{\Psi^{(N)}_i}}^2 \ \delta(E_\tb{k}-E_j-\hbar\omega).
\label{eq:fermi1}}
Here, $\ket{\Psi^{(N)}_{j,\tb{k}}}$ denotes the final state, where the emitted photoelectron 
with momentum \tb{k} and energy $E_\tb{k}$ is typically assumed to be in a scattering state (distorted plane wave, or
time-inverted scattering/LEED state, see e.g. Ref.~\cite{Huefner2003} and references therein). The remaining part of the system is left 
in the excited state $j$ with energy $E_j$ carrying $N-1$ electrons. Both subsystems, the emitted electron and the remaining 
photofragment, are in general still correlated in the combined state $\ket{\Psi^{(N)}_{j,\tb{k}}}$. The wavefunction 
$\ket{\Psi^{(N)}_i}$ represents an initial state
of the many-body system from where the photoelectron will be emitted. The above form of Fermi's Golden rule is strictly valid only
for pure states as initial and final states. In these cases, usually the 
$N$-electron ground state is considered, but also excited eigenstates or superposition states are allowed. For many experimental setups it is not justified
to consider the ground state as initial state for the photoemission process. In particular, in 
pump-probe photoelectron spectroscopy, the system is typically not in the ground state when the photoelectron
is removed from the system.
We illustrate the effect of different initial states for the photoemission process in detail in section \ref{sec:results} and later
in this section.\\
Generally, the coupling element $\hat{\Delta}$ between initial and final states can be written in second 
quantization as
\begin{align}
\label{eq:dipole}
\hat{\Delta}=\sum\limits_{lm,\sigma}\Delta_{lm,\sigma}\hat{c}^\dagger_{l,\sigma}\hat{c}_{m,\sigma},
\end{align}
where $\Delta_{lm,\sigma}=\bra{\varphi^{(1)}_{l,\sigma}}\hat{O}\ket{\varphi^{(1)}_{m,\sigma}}$. 
The coupling to the pump and probe laser pulses is usually considered in the dipole approximation with
length gauge: $\hat{O}  \propto \hat{r}\cdot {\bf E}(r,t)$  or velocity gauge: $\hat{O} \propto \hat{p} \cdot {\bf A}(r,t)$  
(without treating multiphoton processes). ${\bf E}$ and ${\bf A}$ refer to the electric field and the electromagnetic vector potential, respectively.
In the framework of second quantization, the wave functions $\ket{\varphi^{(1)}_{l,\sigma}}$ form a complete set of one-body states. With no external magnetic field applied to the system,
the matrix element is diagonal in spin, since the 
operator does not act on the spin part of the wave function.
The sudden approximation \cite{Almbladh1983,Almbladh2006} allows to decouple the final state
\be{\ket{\Psi^{(N)}_{j,\tb{k}}} \approx \hat{c}^\dagger_\tb{k}\ket{\Psi^{(N-1)}_{j}}\label{eq:sudden}.}
This approximation implies that the final state is a product state between a plane-wave like state
for the emitted electron and the remaining $N-1$ electron many-body state $\ket{\Psi^{(N-1)}_{j}}$.
At this point, we emphasize that the original matrix elements, which contribute to the photocurrent in Eq.~\ref{eq:fermi1}
only contain states with fixed electron number $N$. Therefore, photoemission has to be regarded as a charge neutral
excitation process induced by the presence of a laser field. Only when the excited photoelectron is starting
to spatially separate from the remaining photo-fragment, the system is left in a charged $N-1$ electron state.
This spatial separation is also the basis of the mask approach of Ref.~\cite{DeGiovannini2012}. In general,
the emitted photoelectron can still be entangled with the remaining photo-fragment, for instance in strong coupling situations. However, for weak coupling situation encountered in the range
of the validity of the Fermi's Golden Rule expression for the photocurrent in Eq.~\ref{eq:fermi1}, this entanglement is often neglected.
If no entanglement of the emitted photoelectron and the remaining photofragment remains when the charges
separate, then the state $\ket{\Psi^{(N)}_{j,\tb{k}}}$ can be factorized. This is the basic assumption
of the sudden approximation in Eq.~\ref{eq:sudden}. As a result, in the matrix elements of the photocurrent in Eq.~\ref{eq:fermi1},
the neutral $N$ electron state can be replaced by an ionic $N-1$ electron state. Only in this
sense, we can talk about non-neutral excitations in a photoemission experiment, albeit initially a neutral
excitation has taken place.\\
In terms of the usual fermionic anti-commutation relation, we can write
\begin{align}
\label{eq:anticommutation}
\left\{\hat{c}_{\tb{k}},\hat{c}^\dagger_l\right\}=\delta_{\tb{k}l}\qquad \hat{c}_\tb{k}\hat{c}^\dagger_l \hat{c}_m=\hat{c}_m\delta_{\tb{k}l}-\hat{c}^\dagger_l \hat{c}_m \hat{c}_\tb{k},
\end{align}
and since the state $\ket{\tb{k}}$ of the ejected photoelectron is an energetically high-lying state, 
virtual fluctuations in the reference state can be neglected. Following this argument, 
the last term in Eq. \ref{eq:anticommutation} can be set to zero \cite{Hedin1999} and this allows to write approximately
\begin{align}
\label{eq:anticommutation2}
 \hat{c}_\tb{k}\hat{c}^\dagger_l \hat{c}_m \approx \hat{c}_m\delta_{\tb{k}l}.
\end{align}
Using Eqns.~\ref{eq:dipole} and \ref{eq:sudden}, and the approximation in Eq.~\ref{eq:anticommutation2} to evaluate the matrix 
elements in Eq.~\ref{eq:fermi1}, we arrive at
\ba{\bra{\Psi^{(N)}_{j,\tb{k}}}\hat{\Delta}\ket{\Psi^{(N)}_{i}}&=\sum\limits_{lm,\sigma}\bra{\Psi^{(N-1)}_{j}}\hat{c}_\tb{k}\Delta_{lm,\sigma}\hat{c}^\dagger_{l,\sigma}\hat{c}_{m,\sigma}\ket{\Psi^{(N)}_i}\\
&\approx\sum\limits_{m,\sigma}\Delta_{\tb{k}m,\sigma}\bra{\Psi^{(N-1)}_{j}}\hat{c}_m\ket{\Psi^{(N)}_{i}}.\nonumber
}
In practical applications, the matrix element $\Delta_{lm,\sigma}$ in Eq.~\ref{eq:dipole} is often regarded to be 
constant over the investigated energy range \cite{Huefner2003,Reinert2005}. This assumption is 
only perfectly justified in the high-energy limit (X-ray spectroscopy). In this limit, the photoelectron
spectrum is directly proportional to the spectral function. The sum-over-states  expression for 
the photocurrent in the sudden approximation (SA) is then found to take the form
\be{J^{\rm SA}_\tb{k}(\omega) \approx \frac{2\pi}{\hbar}\sum\limits_{lm,\sigma}\Delta_{\tb{k}l,\sigma}A^{\rm SA}_{lm,\sigma}(E_\tb{k}-\hbar\omega)\Delta_{m,\sigma\tb{k}},
\label{eq:fermi2}} 
where we have introduced the one-body spectral function $A^{\rm SA}_{lm,\sigma}(E_\tb{k}-\hbar\omega)$. In the following, we discuss this quantity for equilibrium and nonequilibrium situations.

\subsection{Spectral function: Sum over states and time-domain formulation}

In this section, we state and define the equilibrium and nonequilibrium spectral function. The derivation of these quantities is given in more detail in the appendix.

\subsubsection{Equilibrium spectral function}

For the present study, it is important to distinguish between equilibrium and nonequilibrium situations. In equilibrium, 
$\ket{\Psi^{(N)}_{i}}$ is a eigenstate of the full vibronic Hamiltonian. The time evolution according to the time-dependent Schr{\" o}dinger equation in Eq.~\ref{eq:td-schroed} is in these cases trivial, since eigenstates are time-invariant up to a phase. In ground-state photoemission spectroscopy, one encounters this situation, if the sample is in its ground-state before it is hit by the photoemission pulse. The equilibrium spectral function is defined as
\begin{align}
\label{eq:spectralfunction}
  A_{lm,\sigma}^{\rm SA}(\omega)= \sum\limits_{j} & \bra{\Psi^{(N)}_0}\hat{c}^\dagger_{l,\sigma}\ket{\Psi^{(N-1)}_j} \\
 & \times\bra{\Psi^{(N-1)}_j}\hat{c}_{m,\sigma}\ket{\Psi^{(N)}_0}\delta(\hbar\omega-E_j). \nonumber
\end{align}
Further, in equilibrium situations only diagonal terms of the spectral function $(l=m)$ need to 
be considered \cite{Stefanucci2013}.\\
Eq.~\ref{eq:spectralfunction} can also be formulated in terms of overlaps of time-evolved states. Using this approach, the calculation of the spectral function
does not rely on a sum-over-states expression. Rather, it can be computed from an explicit time propagation
\ba{\label{eq:spectralfunctionAC}
A_{lm,\sigma}^{SA}(t)=\bra{\tilde{\Psi}^{(N-1)}_{-,l}(t)}{\hat{c}^\dagger_{m,\sigma}}\ket{\Psi^{(N)}_0(t)},
}\\
with the kicked initial state $\ket{\tilde{\Psi}^{(N-1)}_{-,l}(t_0)}={\hat{c}}_l\ket{\Psi_0^{(N)}(t_0)}$. Depending on
the size of the Hilbert space, either Eq.~\ref{eq:spectralfunction} or Eq.~\ref{eq:spectralfunctionAC} are more efficient to evaluate. In our case, we choose 
to directly evaluate Eq.~\ref{eq:spectralfunction} using all eigenstates from our exact diagonalization procedure. However, for larger systems, where a direct diagonalization of the
system Hamiltonian is computationally not feasible anymore, Eq.~\ref{eq:spectralfunctionAC} provides an alternative scheme to obtain the spectral function.\\
A useful relation is the sum rule \cite{Leeuwen2004} that is obeyed by the equilibrium spectral function 
\be{
{S} = \sum_l \int d\omega A_{ll,\sigma}(\omega) = \sum_l \bra{\Psi^{(N)}_0}\hat{c}^{\dagger}_{l,\sigma}\hat{c}^{\,}_{l,\sigma}\ket{\Psi^{(N)}_0},
\label{eq:sumrule}
}
where the value ${S}$ gives the total number of electrons $N$ in the state $\ket{\Psi^{(N)}_0}$. When computing an explicit sum-over-states summation, the limit $ {S} = N$ is only reached, if a complete set of states with a full resolution of
the identity, $\sum_m \ket{\Psi_m^{(N-1)}}\bra{\Psi_m^{(N-1)}} = \mathds{1}$, is inserted in Eq.~\ref{eq:sumrule}. For an incomplete basis of states the sum rule deviates from $N$. Depending on the orthogonality and completeness of the employed states, ${S}$ can then be lower or higher than the total number of electrons in the state $\ket{\Psi^{(N)}_0}$. Therefore, in
practical calculations this sum rule can be exploited to test convergence and the completeness of the employed 
basis set.\\

\subsubsection{Nonequilibrium spectral function}

From the expression for the photocurrent in sudden approximation, Eq.~\ref{eq:fermi2}, the
dependence of the photoelectron spectrum on the reference-state $\ket{\Psi^{(N)}_0}$ becomes 
apparent. As mentioned before, in most cases the system is assumed to be in the ground state. However,
in a pump-probe experiment this assumption is not justified anymore. As we demonstrate in
section \ref{sec:results}, quite sizable changes arise in the photoelectron spectrum when the photoelectron
is ejected from a time-evolving state $\ket{\Psi^{(N)}_0(t)}$ (see discussion in next section).
Ultimately peaks, which were dark for the ground-state as reference state, might become bright 
transitions during time evolution of a vibronic wave packet and can eventually contribute to 
the photoelectron spectrum.\\
In nonequilibrium situations, Fermi's Golden Rule has to be extended to also allow for arbitrary states as reference
states in Eq.~\ref{eq:fermi1}. This can be done straightforwardly in terms of the spectral function and is explicitly calculated in the appendix. Here, we only state the result for the nonequilibrium spectral function
\ba{
\label{eq:spectralfunction_noneq}
  A_{lm,\sigma}^{\rm SA}(t,\omega)= & \sum\limits_{j}  \bra{\Psi^{(N)}_0(t)}\hat{c}^\dagger_{l,\sigma}\ket{\Psi^{(N-1)}_j} \\
 & \times\bra{\Psi^{(N-1)}_j}\hat{c}_{m,\sigma}\ket{\Psi^{(N)}_0(t)}\delta(\hbar\omega-E_j). \nonumber
}
In nonequilibrium situations, the time-evolution of the initial state is nontrivial. Hence, the spectral function expression is not time-invariant and explicitly depends on both the time $t$ and the frequency $\omega$. Physically, we interpret the time $t$ as the delay time between the pump and the probe pulse. As compared to Eq.~\ref{eq:spectralfunction}, we now additionally allow for time-propagated reference states $\ket{\Psi_0^{(N)}(t)}$.\\ 
The nonequilibrium spectral function can also be formulated in the time-domain

\ba{\label{eq:spectralfunctionAC_noneq}
A_{lm,\sigma}^{SA}(t,\tau)=\bra{\tilde{\Psi}^{(N-1)}_{-,l}(t+\tau)}{\hat{c}^\dagger_{m,\sigma}}\ket{\Psi^{(N)}_0(t+\tau,t)}.
}\\
with the kicked initial state $\ket{\tilde{\Psi}^{(N-1)}_{-,l}(t)}={\hat{c}}_l\ket{\Psi_0^{(N)}(t)}$, where the kick with the operator $\hat{c}_l$ acts at time t on the state $\ket{\Psi_0(t)}$.\\
The sum rule of Eq.~\ref{eq:sumrule} also applies in nonequilibrium situations.\\
One point we have to mention is the neglect of the $n$-dependence in the delta function of Eq.~\ref{app:noneq} in the appendix. This approximation allows us to fix peak positions to N-1 electron states. Further, the approximation gives the delta peak position a clear interpretation, rather than the usage of absolute relative energies. Considering the full $n$-dependence in the delta function shifts high energy peaks to lower energy.

\subsection{Approximations for vibronic systems}
The expression for the one-body spectral function in Eq.~\ref{eq:spectralfunction} is still formulated
in terms of correlated energy eigenstates of the full vibronic Hamiltonian. Therefore, the direct evaluation of
this expression is usually a formidable task. For vibronic systems, the most straightforward
approximation is to replace the correlated vibronic initial and final states by factorized 
Born-Oppenheimer states.\\
For the following discussion in section \ref{sec:results}, we define as
single-harmonic approximation (SHA) the case where only the initial state is replaced by the 
corresponding factorized Born-Oppenheimer state in harmonic approximation $\ket{\chi_{00}\phi_0^{(N)}}$
and all final states are retained as correlated vibronic $N-1$ electron states. 
In this case, the spectral function takes the form
\begin{align}
\label{eq:spectralfunctionSHA}
  A_{lm,\sigma}^{\rm SA,SHA}&(\omega)=  \sum\limits_{j} \bra{\chi_{00}\phi_0^{(N)}}\hat{c}^\dagger_{l,\sigma}\ket{\Psi^{(N-1)}_j} \\
 & \times\bra{\Psi^{(N-1)}_j}\hat{c}_{m,\sigma}\ket{\chi_{00}\phi_0^{(N)}}\delta(\hbar\omega-E_j). \nonumber
\end{align}
Interestingly, since the Born-Oppenheimer ground state is by construction not an eigenstate of the full many-body Hamiltonian, already at the level of the SHA the 
expression in Eq.~\ref{eq:spectralfunction_noneq} has to be applied.\\
As further simplification, we can consider a harmonic approximation for both, the involved initial
and final potential energy surfaces and replace the remaining $N-1$ electron states by Born-Oppenheimer 
states in harmonic approximation. This leads to the double-harmonic approximation (DHA) for the
spectral function
\ba{A_{lm,\sigma}^{\rm SA,DHA}(\omega)=\sum\limits_{n,j}&\abs{\braket{\chi_{nj}|\chi_{00}}}^2\bra{\phi_j^{(N-1)}}\hat{c}_{l,\sigma}\ket{\phi_0^{(N)}}\nonumber\\
&\bra{\phi_0^{(N)}}\hat{c}^\dagger_{m,\sigma}\ket{\phi_j^{(N-1)}} \ \delta(\hbar\omega-\epsilon_j)
\label{eq:spectralfunctionDHA}
}\\
In particular, the expression for the DHA shows that the peak-heights
in the photoelectron-spectrum are modulated by Franck-Condon factors. Although this simplifies
practical computations considerably, we show in section \ref{sec:results} that spurious
peaks appear in the DHA of the spectral function, which are not present in the
exact spectral function, Eq.~\ref{eq:spectralfunction}.

\begin{figure*}[t]
  \begin{center}
    \includegraphics[width=\textwidth]{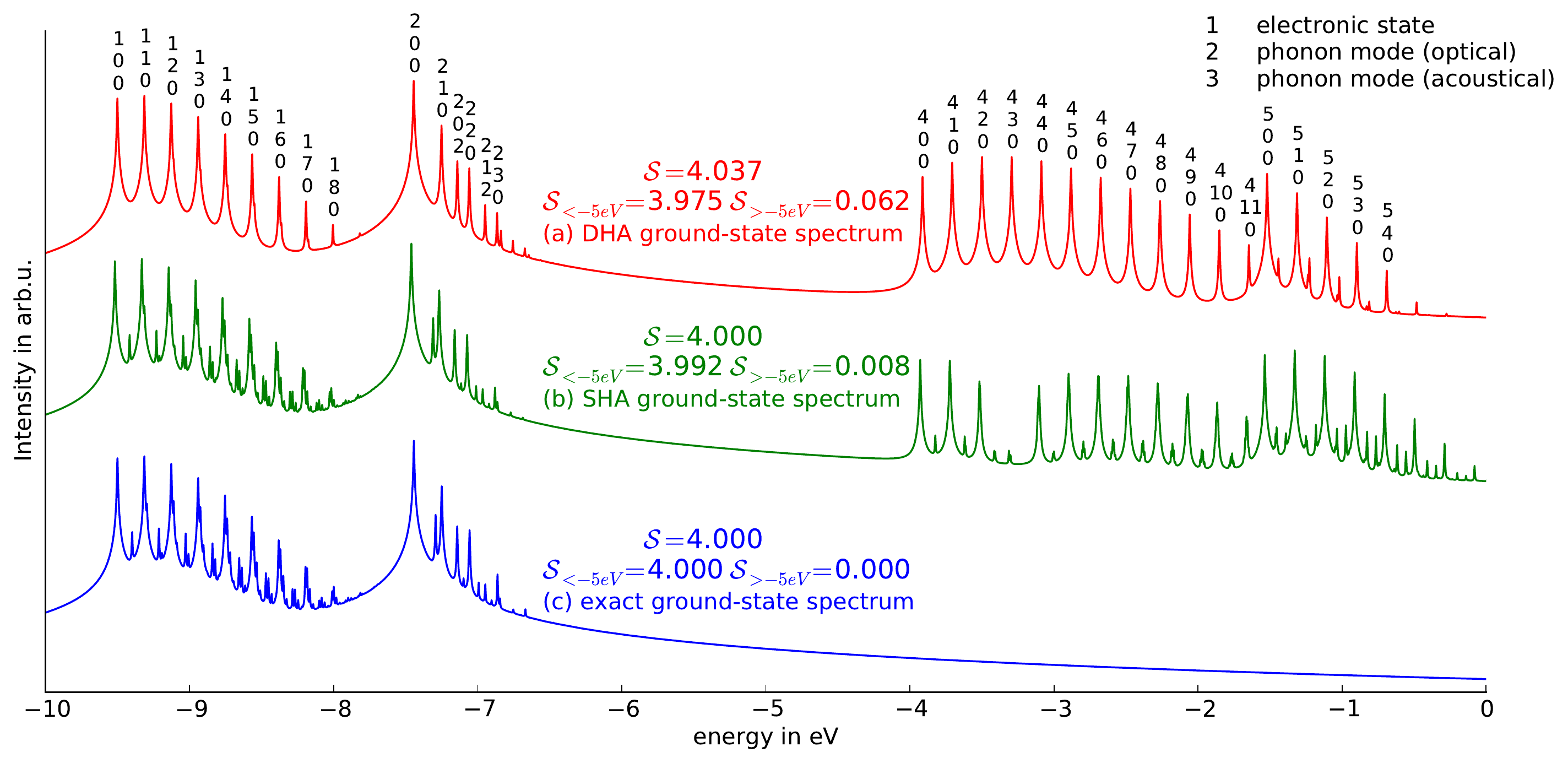}
  \end{center}
  \caption{
    Calculated Photoelectron spectra for Trans-Polyacetylene: (a) The ground-state 
    spectrum in double-harmonic approximation (DHA), (b) the ground-state 
    spectrum in single-harmonic approximation (SHA) (c) the exact ground-state spectrum 
    from the full-quantum calculation. With ${S}$ we refer to the value of the
    sum rule for the spectral function as defined in Eq.~\ref{eq:sumrule}. For a complete set of states this 
    corresponds to the total number of electrons.
    Restricted summations of the sum rule in the energy range above $-5$ eV and
    below $-5$ eV are given by ${S}_{>-5 {\rm eV}}$ and ${S}_{<-5 {\rm eV}}$, respectively. 
    }
  \label{fig:spectra}
\end{figure*}

\section{Results}
\label{sec:results}

\subsection{Comparison of BO and exact ground-state photoelectron spectra}

In this section, we illustrate the different theory levels that we introduced in the
previous section for the calculation of vibronic photoelectron spectra. Due to the dense 
diagonalization that we can perform for our model system of Trans-Polyacetylene, we have 
all correlated states and all required Born-Oppenheimer states available to perform the 
explicit sums over states that arise in the definition of the different spectral functions
in Eqns.~\ref{eq:spectralfunction}, \ref{eq:spectralfunctionSHA}, and \ref{eq:spectralfunctionDHA}.
In the following, we restrict ourselves only to the ground state as initial state for
the photoemission process. We term these spectra ground-state photoelectron spectra.
Later, we lift this restriction to also consider pump pulses and time-evolving reference
states explicitly.

\begin{figure*}[t]
  \begin{center}
    \includegraphics[width=\textwidth]{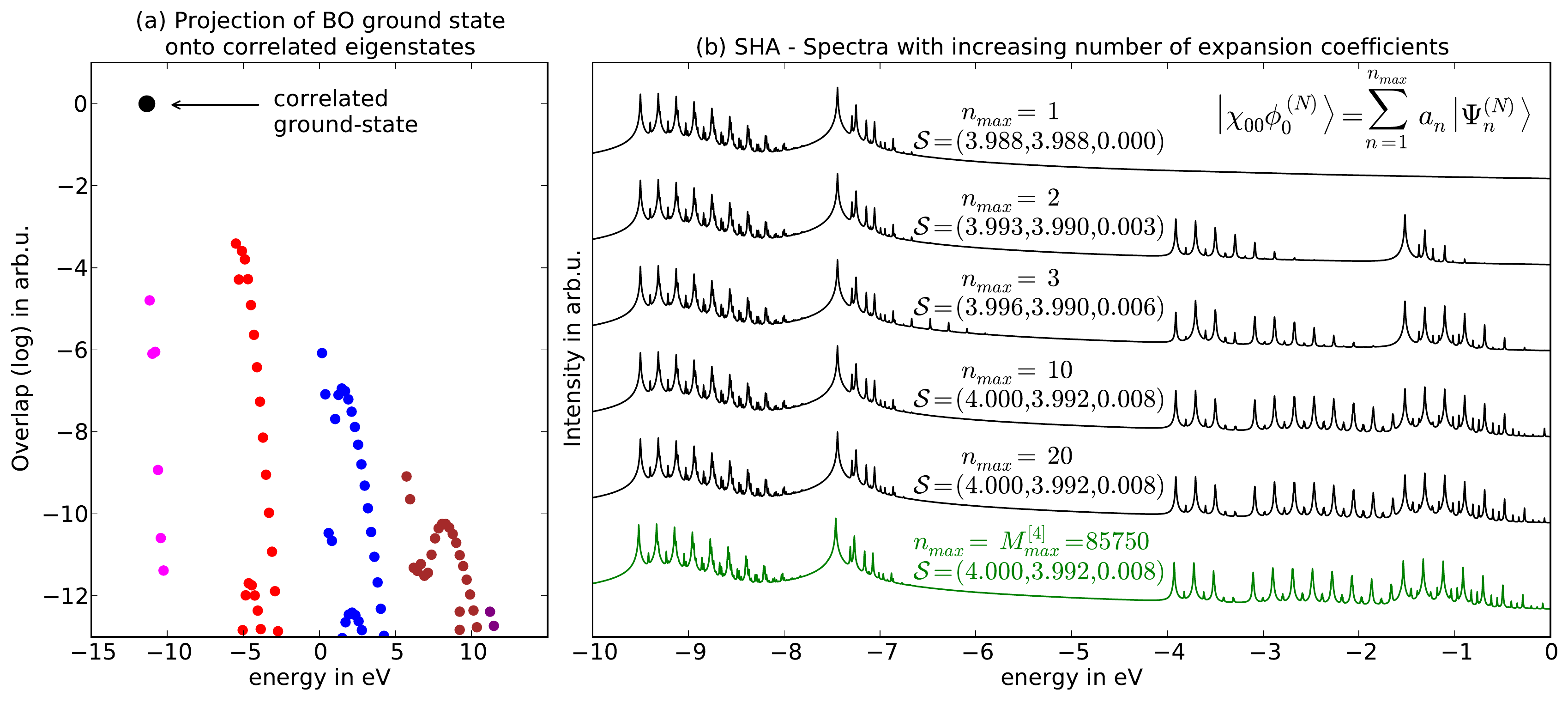}
  \end{center}
  \caption{
    Non-adiabatic contribution to the BO ground state of the Trans-Polyacetylene chain: (a) Projection of the BO ground state onto correlated eigenstates. (b) SHA spectra depending on expansion coefficients. Here, ${S} = ({S}, {S}_{<-5 {\rm eV}}, {S}_{>-5 {\rm eV}})$, as in Fig.~\ref{fig:spectra}.
}
 \label{fig:non-adiabatic}
\end{figure*}

In Fig.~\ref{fig:spectra}, we illustrate spectral functions of the SSH chain 
for three different cases:
In case (a), the spectral function has been calculated in the 
double-harmonic approximation using Eq.~\ref{eq:spectralfunctionDHA}. Spectrum (b)
shows the spectrum calculated in the single-harmonic approximation, where
Eq.~\ref{eq:spectralfunctionSHA} has been employed and in spectrum (c), we show the
exact correlated ground-state spectrum computed from Eq.~\ref{eq:spectralfunction}.
In the figure, the different peaks are labeled according to their corresponding 
quantum numbers (quantum numbers of electronic state, optical mode and acoustical 
mode are shown). In experiment, the spectra are typically plotted as function of the positive binding energy (see e.g. Fig.~4 in Ref.~\cite{Reinert2005}). To connect the plots of the present work to this convention, the absolute value of the x-axis has to be considered to arrive at positive values for the binding energy. Furthermore, for all spectra a Lorentzian broadening of the form
\be{f(E;E_0,\gamma)=\frac{1}{\pi}\frac{\gamma}{\left(E-E_0\right)^2+\gamma^2}.}
with $\gamma$=0.002 eV has been used. Note, that a conventional broadening of
0.1 eV that is employed frequently for purely electronic Green's functions would completely
wash out the vibrational side bands. To resolve here the vibrational side-bands of
the photoelectron spectrum a much smaller broadening of 0.002 eV has to be employed.
Note, that this broadening is also about an order of magnitude smaller than $1/40$ eV,
which gives a typical energy scale for vibronic motion at room temperature. In experiment, the vibrational
sidebands are hence only clearly visible in a low temperature limit.\\
In comparison to the exact spectrum in (c), we conclude from Fig.~\ref{fig:spectra} that 
DHA and SHA, both accurately
predict the peak positions corresponding to the optical phonon mode in the energy range of the spectrum from -10 eV to -5 eV, but 
the spectra reveal clear differences in the energy range from -5 eV up to 0 eV. The accurate location
of the peaks is in accord with the quality of the approximate energy values shown 
in Tab.~\ref{tab:overlap}. On the other hand, 
peak heights in the DHA are not accurate: peaks, which correspond to the optical phonon mode 
are most dominant in the spectrum and their broadening overlaps and even hides peaks, which 
correspond to mixed or acoustical phonon modes. \\
As additional information, we also show in Fig.~\ref{fig:spectra} the sum rule calculated with Eq.~\ref{eq:sumrule} for
each spectrum. The DHA spectrum violates the sum rule due to the non-completeness 
of the approximation, as discussed in Sec.~\ref{sec:spectral}. In all three spectra, most of the spectral amplitude is located in energy 
areas below -5 eV, while only less than two percent of the spectral weight 
is located in the energy range above -5 eV in the DHA and SHA spectra.
The most prominent feature between the different spectra is that in the DHA and SHA
spectra spurious peaks appear above -5 eV that are not present in the exact correlated ground-state spectrum.
We discuss the origin of this artefact of the DHA and SHA in detail 
in the next section.

\begin{figure*}[t]
\begin{center}
    \includegraphics[width=\textwidth]{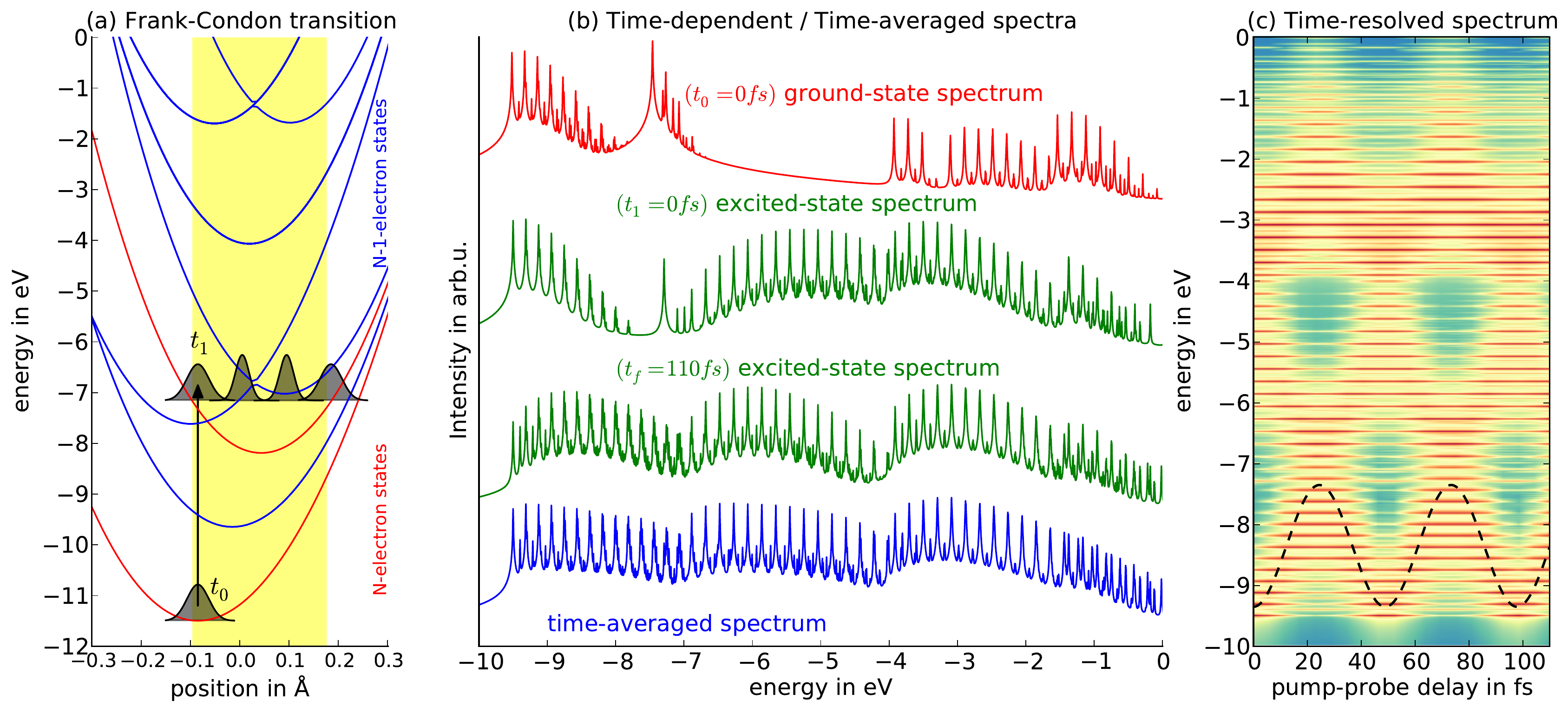}
\end{center}
  \caption{
    (a) Illustration of Frank-Condon transition: The excited $N$-electron initial state propagates on the
    Born-Oppenheimer surface of the first excited electronic state. $N$-electron 
    potential-energy surfaces are shown in red, $N-1$-electron potential-energy surfaces in blue. The 
    oscillation spread of the center of the wavepacket is indicated by a yellow background. The different 
    wavepacket shapes indicate the squeezing of the vibronic state. (b) Spectra 
    at different time-steps (the first spectrum at $t_0$ corresponds to spectrum (b) in Fig.~\ref{fig:spectra}), 
    (c) All obtained spectra plotted time-resolved, the color code refers to high intensity in red and low intensity in blue color. The dashed
    black line shows the motion of the center of the nuclear wavepacket as function of pump-probe delay. The color code refers to high intensity in red and low intensity in blue color.
    }  \label{fig:time_res} 
\end{figure*}

\subsection{Non-adiabaticity in ground-state photoelectron spectroscopy}

The prominent differences in the energy range from -5 eV to {0 eV} between 
the DHA spectrum and the exact correlated spectrum shown in Fig.~\ref{fig:spectra} 
have two equally important contributions: As indicated by the name {\em{double}}-harmonic 
approximation, one performs two harmonic approximations in the DHA. It turns out that both 
harmonic approximations contribute independently to the spurious peaks in the 
spectrum. We can isolate the effect of each of the two harmonic approximations 
by comparing to the single-harmonic approximation. Since according to our 
definition in Eq.~\ref{eq:spectralfunctionSHA}, we use correlated final states 
in the SHA, the only remaining approximation in the SHA is the factorized and harmonic Born-Oppenheimer 
initial state. Comparing the sum rules for the spectral function of the DHA 
with the spectral function of the SHA in Fig.~\ref{fig:spectra} for the upper 
part of the spectrum ({-5 eV} to 0 eV) shows that the spectral weight of the 
spurious peaks is reduced from 1.6\% in DHA to 0.2\% in SHA. The remaining 
spurious amplitude, and hence the differences between the SHA spectrum in Fig.~\ref{fig:spectra} (b) and 
the exact spectrum in Fig.~\ref{fig:spectra} (c), is caused by the factorized Born-Oppenheimer initial 
state in the SHA.\\
To illustrate this further, we expand the Born-Oppenheimer ground state in the complete set of correlated eigenstates 
of the full many-body Hamiltonian from Eq.~\ref{eq:FullStaticSE}
\ba{
\label{eq:BOexpansion}
\ket{\chi_{00} \, \phi_0^{(N)}} & = \sum_{n=1}^{n_{\rm max}} \brakete{\Psi_n^{(N)}}{\chi_{00} \, \phi_0^{(N)}}\ket{\Psi_n^{(N)}} \\
& = \sum_{n=1}^{n_{\rm max}} a_n\ket{\Psi_n^{(N)}} \nonumber
.} 
The magnitude for the different expansion coefficients $a_n$ is shown in Fig.~\ref{fig:non-adiabatic} (a)
in logarithmic scale. As expected, 
the highest overlap is found between the Born-Oppenheimer ground state and the exact 
correlated ground state. For the present system, this overlap is equal to 0.9986 (see Tab.~\ref{tab:overlap}) 
and is marked as a black dot in the graph. The following corrections are orders of 
magnitudes smaller. In Fig.~\ref{fig:non-adiabatic} (a), we illustrate with different colors
the overlaps $a_n$ for $n_{\rm max} \ge 1 $ with magnitude larger than $10^{-13}$. The overlaps can be grouped in
different sets, which allow to identify different PES in terms of Fig.~\ref{fig:cartoon-dha}.
In Fig.~\ref{fig:non-adiabatic} (b), we show the SHA spectral function for different upper limits of summation $n_{\rm max}$ 
in the expansion of the Born-Oppenheimer initial state. If only the coefficient $a_1 = 0.9986$ 
with the highest overlap is included, we recover the exact correlated ground-state spectrum.
This is shown in the upper spectrum in Fig.~\ref{fig:non-adiabatic} (b).
Note, that we are not renormalizing the state in Eq.~\ref{eq:BOexpansion} after truncation,
so that the sum rule corresponds for $n_{\rm max} =1 $ to ${S} = 4 \cdot a_1^2  = 3.9888$.
When more and more expansion coefficients $a_n$ with $n>1$ are included in the expansion, 
the artificial peaks shown in Fig.~\ref{fig:spectra} in the range from {-5 eV} to 0 eV start to emerge. This is illustrated in the 
sequence of spectra in Fig.~\ref{fig:non-adiabatic} (b). 
When the expansion of Eq.~\ref{eq:BOexpansion} is inserted 
in Eq.~\ref{eq:spectralfunctionSHA}, the spurious peaks arise due to additional cross
and diagonal terms in the spectral function, which involve excited correlated eigenstates. Hence, we conclude that 
the artificial peaks are due to the factorized nature of the Born-Oppenheimer ground state. We
emphasize, that the spurious spectral weight appears, both for the Born-Oppenheimer ground state
in harmonic approximation, as well as for the exact Born-Oppenheimer ground state without the harmonic approximation. In both cases,
the expansion in Eq.~\ref{eq:BOexpansion} in terms of correlated vibronic eigenstates has in 
general more than one term ($n_{\rm max} > 1 $) and hence additional cross and diagonal terms in the spectral function necessarily appear.
As we have demonstrated in Tab.~\ref{tab:overlap}, for the present model of Trans-Polyacetylene 
the overlap between exact Born-Oppenheimer, harmonic Born-Oppenheimer, and exact correlated ground state
is very high due to the rather harmonic nature of the Su-Schrieffer-Heeger model. Nevertheless, the spurious spectral
weights already have a magnitude of about 1.6\% in DHA. For any molecular system, which is
less harmonic than our model, a larger contribution to the spurious spectral peaks is expected,
since in the expansion more terms with a larger weight of expansion coefficients $a_n$ contribute.
In this sense, the present system can be regarded as best-case scenario and in general the spurious spectral
peaks are more pronounced. However, in the limit of large nuclear masses, the Born-Oppenheimer approximation
becomes more accurate. In this limit, the Born-Oppenheimer ground state of the system becomes identical to the
correlated ground state, hence leading to identical spectra.\\
One way to correctly incorporate nonadiabatic effects could be the inclusion of non-adiabatic
couplings in the Born-Huang expansion (Eq.~\ref{Born-Huang}). Other alternatives could rely on an explicitely correlated
ansatz for the combined electron nuclear wavefunction, as e.g. in an electron-nuclear coupled
cluster approach \cite{Ko2011}, or in a multi-component density functional theory approach
for electrons and nuclei \cite{Kreibich2008}.

\subsection{Time-resolved pump-probe photoelectron spectra}
\label{sec:prop_fc}

So far, we have considered the ground state as the reference state for the calculation
of the spectral function. In this section, we turn our attention to explicitly 
time-resolved vibronic photoelectron spectra. All calculations for the remaining part
of the paper are done with the exact Hamiltonian and are based on the exact time-evolution
of the correlated time-dependent Schr\"odinger equation. 
To illustrate pump-probe photoelectron spectra for vibronic systems, we consider two 
different examples. In example (1), we initially excite our system with a Franck-Condon 
transition, while in example (2), we explicitly include a short femtosecond laser pulse 
with Gaussian envelope in our real-time propagations to simulate the pump pulse. 
We start in the present section with example (1).
\subsubsection{Time-resolved photoelectron spectra with initial Frank-Condon excitation}
In our first example, we excite the SSH chain from the Born-Oppenheimer ground state
to the first
excited charge neutral $N$ electron state $\phi_1^{(N)}$, while the vibrational state remains
in the ground state configuration $\chi_{00}$. After excitation, the initial state for the time propagation is still a factorized Born-Oppenheimer state of the form
\ba{\label{eq:FCinitial_state} \ket{\Psi^{(N)}(t = t_0)} = \ket{\chi_{00} \, \phi_1^{(N)}}.}
This type of Franck-Condon transition takes here the role of the pump pulse and is
illustrated in Fig.~\ref{fig:time_res} (a) in the left panel.
The initial state in Eq.~\ref{eq:FCinitial_state} is then propagated in real-time 
with the full {\em correlated} Hamiltonian in the combined electronic and 
vibrational Fock space of the model. Since the factorized excited Franck-Condon state 
is not an eigenstate of the correlated many-body Hamiltonian, a wave-packet
propagation is launched with this initial state, which resembles predominantly the motion of a Born-Oppenheimer
state in the first excited potential energy surface. We propagate from $t_0=0$ fs to a final
time of $t_f=110$ fs, which corresponds to about $9/4$ of the oscillation period of the nuclear wavepacket 
in the excited state. The oscillation spread of the center of the nuclear wavepacket is indicated by a yellow 
background in Fig.~\ref{fig:time_res} (a).\\
After a certain delay time $\tau$ we simulate a probe pulse by recording the photoelectron spectrum 
in terms of the spectral function. This amounts to replacing the reference state $\ket{\Psi^{(N)}_0}$ 
in Eq.~\ref{eq:spectralfunction} with the time-evolved state $ \ket{\Psi^{(N)}(\tau)}$ at the 
pump-probe delay time $\tau$
\begin{align}
\label{eq:TDspectralfunction}
  A_{lm,\sigma}^{\rm SA}(\tau,\omega)= &\sum\limits_{j}  \bra{\Psi^{(N)}(\tau)}\hat{c}^\dagger_{l,\sigma}\ket{\Psi^{(N-1)}_j} \\
 & \times\bra{\Psi^{(N-1)}_j}\hat{c}_{m,\sigma}\ket{\Psi^{(N)}(\tau)}\delta(\hbar\omega-\epsilon_j). \nonumber
\end{align}
\begin{figure*}[t]
\begin{center}
    \includegraphics[width=\textwidth]{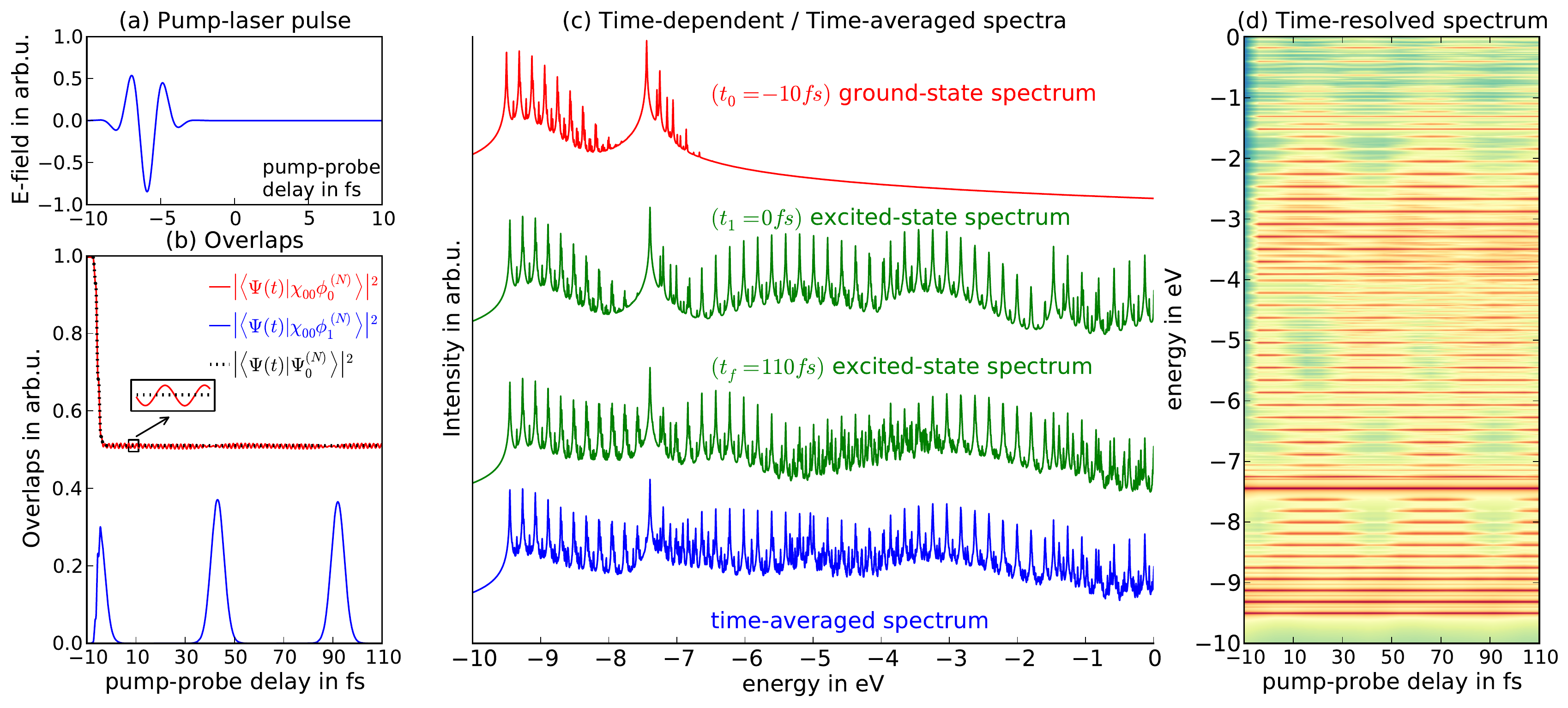}
\end{center}
  \caption{
    Time-dependent pump-probe photoelectron spectroscopy: (a) Amplitude of laser-pulse
    as given in Eq.~\ref{eq:dipole}, (b) overlaps of time-evolved state with electronic ground and first 
    excited Born-Oppenheimer states and the exact correlated ground state, (c) spectra 
    at different time-steps (the first spectrum at $t_0$ corresponds to spectrum (c) in Fig.~\ref{fig:spectra}), 
    (d) all obtained spectra plotted time-resolved, the color code refers to high intensity in red and low intensity in blue color.
    }  \label{fig:pump_pulse} 
\end{figure*}
In Fig.~\ref{fig:time_res} (b), the corresponding photoelectron spectra 
for two different delay times of $\tau = t_1 = 0$ fs and $\tau = t_f = 110$ fs is shown in green color. The spectra after 
different pump-probe delays show that several peaks gain spectral amplitude, which were dark
in the ground-state spectrum and conversely other peaks loose amplitude, which were bright before. 
A more complete picture of the underlying wavepacket dynamics can be obtained by plotting the 
spectral function $A_{lm,\sigma}^{\rm SA}(\tau,\omega)$ as continuous function of the delay 
time $\tau$. This is shown in Fig.~\ref{fig:time_res} (c). Here, every slice of the 2D plot at fixed 
$\tau$ corresponds to one recorded spectrum. The color code indicates the intensity of the peaks, 
with red color for high photoelectron amplitude and blue color for lower amplitude. The spacing
between neighboring peaks corresponds to different vibronic states in the same potential-energy surface.
Besides the spectral function, we also plot with a dashed line in Fig.~\ref{fig:time_res} (c) the 
center of the nuclear wavepacket (first moment)
as function of the delay time $\tau$. The oscillation time $T_0$ is in this case $T_0=48.94$ fs.
The 2D plot of the spectral function nicely illustrates that the gain and
loss of spectral amplitude as function of pump-probe delay time $\tau$
is directly linked to the underlying 
nuclear wavepacket motion. This is similar to optical pump-probe spectroscopy, which provides
a stroboscopic picture of the nuclear dynamics of the system. The notable difference here is
that we record outgoing photoelectrons, and therefore states, which have vanishing optical 
matrix elements with the initial state can also be monitored.\\
Typically, in non-time-resolved pump-probe photoelectron experiments, time-averages of 
spectra are recorded. We therefore include in Fig.~\ref{fig:time_res} (b) in blue color 
also a time averaged spectrum that is computed according to
\begin{align}
 A_{lm,\sigma}^{\rm SA}(\omega) = \frac{1}{t_f - t_1} \int_{t_1}^{t_f} A_{lm,\sigma}^{\rm SA}(\tau,\omega) d\tau
\end{align}
and that can be viewed as an average of the 2D contour data of Fig.~\ref{fig:time_res} (c)
along the axis of the delay time $\tau$. The average spectrum is useful in determining
in which spectral regions the emitted photoelectrons can be found over certain oscillation
periods.

\subsubsection{Time-resolved photoelectron spectra with explicit pump pulse}

In our second example, we investigate an explicit simulation of a pump-probe experiment in
real-time.
Compared to the Franck-Condon excitation, which was based only on the selection of
a specific excited initial state, a more appropriate description of the excitation of
the system can be realized by explicitly including the pump pulse into the time-propagation. 
For the following discussion, we therefore add a dipole-coupling term to the Hamiltonian in 
Eq.~\ref{eq:ssh1}
\ba{\hat{H}_{ssh}(t)=&\hat{H}_\pi+\hat{H}_{ph}+\hat{H}_{\pi-ph} \label{eq:dipole-coup} \\
&+ \hat{H}_{\pi,E}(t) + \hat{H}_{ph,E}(t),\nonumber\\
\hat{H}_{\pi,E}(t)=& -e\sum\limits_{n,\sigma}\,x_n\,\hat{c}^\dagger_{n,\sigma}\hat{c}_{n,\sigma} \cdot E(t), \nonumber\\
\hat{H}_{ph,E}(t)=&\sum\limits_{n}q_n\,\hat{u}_{n} \cdot E(t), \nonumber\\
E(t) =& E_0 \, \text{exp}\left(-\left(t-t_0\right)^2/\sigma^2\right) \, \sin{\omega_l t}.\nonumber
}
Here, $x_n$ refers to the real-space position of site n, $e$ to the elementary electric charge and $q_n$ 
to the charge of the nuclei n (in the present case, we choose $q_n = e$). Note, that the laser pulse
couples to both, to the dipole moment of the electrons and to the nuclear dipole. For the 
electric field of the 
pump pulse $E(t)$ we use a Gaussian envelope with midpoint $t_0= -6$ fs, maximum envelope 
$E_0$ = 0.85 $V/$\AA~and variance $\sigma = 1.5$ fs. As 
carrier wave, we choose a sine function with frequency $\omega_l = \Delta E/\hbar=6.20$ fs$^{-1}$. 
The frequency of the laser pulse is chosen to be resonant for a Frank-Condon like transition with $\Delta E$
that corresponds to the example in Sec.~\ref{sec:prop_fc} (1). 
For the time-propagation with the time-dependent Hamiltonian $\hat{H}_{ssh}(t)$, we use as
before a Lanczos propagator, but in addition we employ 
an exponential midpoint scheme \cite{Castro2004} to account for the time-dependence of the Hamiltonian.
For the propagation, we choose the exact correlated vibronic ground state as initial state. This state is then 
propagated with fully correlated many-body Hamiltonian including the dipole coupling to the pump
laser as given in Eq.~\ref{eq:dipole}.\\
In Fig.~\ref{fig:pump_pulse} (a), we show the amplitude of the external laser pulse. The pulse 
starts at $t_0=-10$ fs and is switched off at $t_1=0$ fs. As the state evolves, we compute the
overlaps with the Born-Oppenheimer states as function of time. In Fig.~\ref{fig:pump_pulse} (b)
we show, which states are populated during the propagation. While at the initial time almost the full
population is in the BO ground state ($\ket{\chi_{00} \, \phi_0^{(N)}}$), the population moves 
within the first 10 fs to the state $\ket{\chi_{00} \, \phi_1^{(N)}}$, i.e. the state, which 
corresponds to a Frank-Condon transition out of state $\ket{\chi_{00} \, \phi_0^{(N)}}$. After 
the first initial population of $\ket{\chi_{00} \, \phi_1^{(N)}}$, the populations indicate two 
competing processes: First, the pump laser pulse transfers population from $\ket{\chi_{00} \, \phi_0^{(N)}}$ 
to $\ket{\chi_{00} \, \phi_1^{(N)}}$, since this transition is resonant. Second, once population 
occurs in $\ket{\chi_{00} \, \phi_1^{(N)}}$, this population induces a wavepacket motion on the 
first excited potential-energy surface, as seen in Sec.~\ref{sec:prop_fc} (1). Therefore, the 
system never reaches a situation, where only two states are present in the system $\ket{\chi_{00} \, \phi_0^{(N)}}$ and $\ket{\chi_{00} \, \phi_1^{(N)}}$, as
it would be in a Franck-Condon picture of the excitation process. After the end of the pump pulse at $t=0$ fs, the projection of the time-evolving
state on the correlated vibronic ground state is constant in time and the ground state maintains a population of about 50\%. In contrast, the projection of the correlated time-evolving state on the BO ground state exhibits small oscillations which is shown
in the inset of Fig.~(b) and which arise due to the small deviations between the BO ground state and the exact correlated ground state.
As in the example before, we record photoelectron spectra as function of pump and probe
delay. The photoelectron spectrum at time $t=t_0=-10$ fs is given in red color in Fig.~\ref{fig:pump_pulse} (c).
This spectrum is identical to the spectrum also shown in Fig.~\ref{fig:spectra} (c). 
The spectra after the pulse has been switched off are shown in green color and correspond
to delay times of $t=0$ fs and $t=110$ fs.
As before, we also show the time-averaged spectrum in blue color. In contrast to the Franck-Condon excitation
in example (1), the photoelectron spectra in this case show a pronounced large peak at about $-7.5$ eV. This peak
arises due to the remaining population of the Born-Oppenheimer ground state $\ket{\chi_{00} \, \phi_0^{(N)}}$.
In Fig.~\ref{fig:pump_pulse} (d), we show similar to Fig.~\ref{fig:time_res} (c) the spectral function
$A_{lm,\sigma}^{\rm SA}(\tau,\omega)$ as function of the pump-probe delay $\tau$. As before in the Franck-Condon
case, also with an explicit pump pulse, we can trace the nuclear wavepacket dynamics in the photoelectron
spectrum. Overall, the time-evolution of the Franck-Condon excitation captures large parts of the exact vibronic 
spectrum of example (2). The notable differences in the explicitly time-dependent and fully correlated vibronic case in Fig.~\ref{fig:time_res} (c) 
are the additional population of the Born-Oppenheimer ground state which maintains a large spectral weight 
at about $-7.5$ eV also for different delay times and some small non-adiabatic contributions to the spectrum 
in the energy range from $-4$ eV to $-2$ eV.

\section{Conclusion}
\label{sec:conc}

In summary, we have analyzed and quantified non-adiabatic contributions
to the equilibrium and nonequilibrium photoelectron spectra in a model 
system for Trans-Polyacetylene.
We find that for low-lying states the harmonic Born-Oppenheimer 
photoelectron spectrum acquires in comparison to the exact photoelectron
spectrum spurious spectral weight, which also
persists when either the initial state of the photoemission process
or the final state is replaced by correlated vibronic states. The origin
of this behavior can be traced back to the factorized nature of the
involved initial or final Born-Oppenheimer states. Only when both, 
initial and final photoemission states, are taken as correlated vibronic
states the spurious spectral peaks are suppressed. We analyze this
in detail by expanding the Born-Oppenheimer ground state in the complete
set of correlated vibronic eigenstates of the full Hamiltonian. Inserting
this expansion into the equilibrium form of the spectral function shows
that additional cross and diagonal terms, which involve excited correlated eigenstates,
are responsible for the spurious spectral weight.\\
For the example of an initial Franck-Condon transition and for an
explicit pump-pulse excitation we have demonstrated with explicit
real-time propagations of the coupled Polyacetylene chain how the
vibronic wavepacket evolution can be traced in the photoelectron
spectrum as function of pump-probe delay. \\
Prospects for future work include the study of temperature and
pressure dependence of the photoelectron spectra as well as an
extension of the present femtosecond laser excitation to 
ultrafast photoelectron spectroscopy with attosecond laser pulses in real nanostructured
and extended systems. Another line of research is linked to the development of xc functionals
for TDDFT capturing the effects discussed in this work, e.g. based on electron-nuclear multicomponent
density functional theory \cite{Kreibich2008}.

\begin{acknowledgments}
The authors thank Professor Matthias Scheffler for his support and Professor 
Ignacio Franco for useful discussions during the preparation of the manuscript.\\
We acknowledge financial support from the European Research Council Advanced
Grant DYNamo (ERC-2010-AdG-267374), Spanish Grant (FIS2010-21282-C02-01),
Grupos Consolidados UPV/EHU del Gobierno Vasco (IT578-13), Ikerbasque and the
European Commission projects CRONOS (Grant number 280879-2 CRONOS CP-FP7).
\end{acknowledgments}



\appendix*

\section{Appendix}

\subsection{Spectral function}
\label{app:section_general}

The one-body spectral function is defined as follows \cite{Stefanucci2013}:
\ba{\label{app:spectral_general}
A_{ij}(t,t') &= \bra{\Psi_0}\left\{\hat{c}_i(t)\hat{c}^{\dagger}_j(t')\right\}\ket{\Psi_0}\\
A_{ij}(t,t') &= \bra{\Psi_0}\hat{c}_j^{\dagger}(t')\hat{c}_i(t) + \hat{c}_i(t)\hat{c}^{\dagger}_j(t') \ket{\Psi_0}\nonumber\\
A_{ij}(t,t') &= A^{<}_{ij}(t,t') + A^{>}_{ij}(t,t')\nonumber
,}
with $t'>t$. The operators $\hat{c}$ and $\hat{c}^\dagger$ are here written in the Heisenberg picture. The index refers to a combined index $i = (n,\sigma)$ and $j = (m,\sigma^{\prime})$, where $n$ and $m$ refer to the site number and $\sigma$ and $\sigma^\prime$ refer to spin up or spin down.\\ 
In this work, we only consider the first part of the commutator $A^{<}_{ij}(t,t')$, since we are only interested in photoemission spectra. The second term $A^{>}_{ij}(t,t')$ leads to inverse photoemission spectra \cite{Stefanucci2013}. In the following discussion, we distinguish two cases: 1. if $\ket{\Psi_0}$ is an eigenstate of the system Hamiltonian, we work in an equilibrium framework, 2. if $\ket{\Psi_0}$ is not an eigenstate of the system Hamiltonian, we have to work in a nonequilibrium framework.

\subsection{Equilibrium spectral function}
\label{app:section_eq}

The equilibrium spectral function applies for situations, where
$\ket{\Psi_0}$ is an eigenstate
of the corresponding many-body Hamiltonian $H$ of the system. Hence, we can write Eq.~\ref{app:spectral_general} in terms of an time-correlation function
\ba{
A_{ij}(t,t') &= \bra{\Psi_0}\hat{c}^\dagger_j(t')\hat{c}_i(t)\ket{\Psi_0}\\
&=\bra{\Psi_0}e^{iHt'/\hbar}\hat{c}^\dagger_j e^{-iHt'/\hbar}e^{iHt/\hbar}\hat{c}_ie^{-iHt/\hbar}\ket{\Psi_0}\nonumber\\
&=\bra{\Psi_0}\hat{c}^\dagger_j e^{-iH(t'-t)/\hbar} \hat{c}_i\ket{\Psi_0}e^{i E_0(t'-t)/\hbar}\nonumber\\
&=\bra{\Psi_0(\tau)}\hat{c}^\dagger_j \ket{\tilde{\Psi}(\tau)}\nonumber 
}
with $\tau = t'-t$ and the initial condition $\ket{\tilde{\Psi}(\tau = \tau_0)}=\hat{c}_i\ket{\Psi_0}$. \\
Eq.~\ref{app:spectral_general} can be reformulated to get a sum-over-states expression. This is accomplished by the insertion of an complete set of states $\sum\limits_m \ket{\Psi_m}\bra{\Psi_m} = \mathds{1}$ and a Fourier transform with respect to the time-difference $\tau= t'-t$
\ba{
A^{<}_{ij}(t,t') &= \bra{\Psi_0}\hat{c}^\dagger_j(t')\hat{c}_i(t)\ket{\Psi_0}\\
=&\bra{\Psi_0}e^{iHt'/\hbar}\hat{c}^\dagger_j e^{-iHt'/\hbar}e^{iHt/\hbar}\hat{c}_ie^{-iHt/\hbar}\ket{\Psi_0}\nonumber\\
=&\sum\limits_m\bra{\Psi_0}\hat{c}^\dagger_j\ket{\Psi_m}\bra{\Psi_m}\hat{c}_i\ket{\Psi_0}e^{i (E_0-E_m)(t'-t)/\hbar}\nonumber\\
A^{<}_{ij}(\omega)=&\int\limits^\infty_{-\infty}\frac{d\tau}{{2\pi}} \sum\limits_m\bra{\Psi_0}\hat{c}^\dagger_j\ket{\Psi_m}\bra{\Psi_m}\hat{c}_i\ket{\Psi_0}\nonumber\\
&\times e^{i (E_0-E_m-\hbar\omega)\tau/\hbar}\nonumber\\
=&\sum\limits_m\bra{\Psi_0}\hat{c}^\dagger_j\ket{\Psi_m}\bra{\Psi_m}\hat{c}_i\ket{\Psi_0}\delta\left(E_0-E_m-\hbar\omega\right)\nonumber\\
A^{<}_{ii}(\omega)=&\sum\limits_m\abs{\bra{\Psi_0}\hat{c}^\dagger_i\ket{\Psi_m}}^2\delta\left(E_0-E_m-\hbar\omega\right)\nonumber
}

\subsection{Nonequilibrium spectral function}
\label{app:section_noneq}

In nonequilibrium situations, $\ket{\Psi_0}$ is not an eigenstate of the many-body Hamiltonian $H$. Nevertheless, it is also possible to formulate the spectral function in Eq.~\ref{app:spectral_general} as time-correlation function involving propagated states
\ba{
A^{<}_{ij}(t,t') &= \bra{\Psi_0}\hat{c}^\dagger_j(t')\hat{c}_i(t)\ket{\Psi_0}\\
&=\bra{\Psi_0}e^{iHt'/\hbar}\hat{c}^\dagger_j e^{-iHt'/\hbar}e^{iHt/\hbar}\hat{c}_ie^{-iHt/\hbar}\ket{\Psi_0}\nonumber\\
&=\bra{\Psi_0(t')}\hat{c}^\dagger_j e^{-iH(t'-t)/\hbar} \hat{c}_i\ket{\Psi_0(t)}\nonumber\\
A^{<}_{ij}(t,\tau)&=\bra{\Psi_0(\tau+t)}\hat{c}^\dagger_j \ket{\tilde{\Psi}(\tau+t,t)}\nonumber
.}
We introduce the relative time $\tau=t'-t$, as in Sec.~\ref{app:section_eq}, while t keeps its initial denotation. The state $\ket{\tilde{\Psi}(\tau+t,t)}$ is defined as  $\ket{\tilde{\Psi}(\tau+t,t)}= e^{-iH\tau/\hbar} \hat{c}_i\ket{\Psi_0(t)}$, meaning the kick $\hat{c}_i$ on the wavefunction acts at time t during the time propagation. A Fourier transform with respect to the relative time $\tau$ yields the general expression for the sum-over-states expression
\ba{\label{app:noneq}
A^{<}_{ij}(t,t') =& \bra{\Psi_0}\hat{c}^\dagger_j(t')\hat{c}_i(t)\ket{\Psi_0}\\
A^{<}_{ij}(t,\tau)=&\bra{\Psi_0}e^{iH (\tau+t)/\hbar}\hat{c}^\dagger_j e^{-iH\tau/\hbar}\hat{c}_ie^{-iHt/\hbar}\ket{\Psi_0}\nonumber\\
=&\sum\limits_{n,n',m}e^{i\tau/\hbar \left(E_{n'}-E_{m}\right) +it/\hbar \left(E_{n'}-E_n \right)}\nonumber\\
&\times \brakete{\Psi_0}{\Psi_{n'}}\bra{\Psi_{n'}}\hat{c}^\dagger_j \ket{\Psi_m}\bra{\Psi_m}\hat{c}_i\ket{\Psi_n}\brakete{\Psi_n}{\Psi_0}\nonumber\\
A^{<}_{ij}(t,\omega) =&\int\limits^\infty_{-\infty}\frac{d \tau}{{2\pi}} \sum\limits_{n,n',m}e^{i\tau/\hbar\left(E_{n'}-E_{m}-\hbar\omega\right)+it/\hbar \left(E_{n'}-E_n \right)
}\nonumber\\
&\times\brakete{\Psi_0}{\Psi_{n'}}\bra{\Psi_{n'}}\hat{c}^\dagger_j \ket{\Psi_m}\bra{\Psi_m}\hat{c}_i\ket{\Psi_n}\brakete{\Psi_n}{\Psi_0}\nonumber\\
=&\sum\limits_{n,n',m}e^{it/\hbar \left(E_{n'}-E_{n}\right)} \delta\left(E_{n'}-E_m-\hbar\omega \right)\nonumber\\
&\times \brakete{\Psi_0}{\Psi_{n'}}\bra{\Psi_{n'}}\hat{c}^\dagger_j \ket{\Psi_m}\bra{\Psi_m}\hat{c}_i\ket{\Psi_n}\brakete{\Psi_n}{\Psi_0} \nonumber
}
In our simulations, we neglect the energy dependence of the delta function in the last equation. Hence, we replace the term
$E_{n'}$ by the energy $E_0$ of the state $\Psi_0$. This leads to
\ba{
A^{<}_{ij}(t,\omega)& =\sum\limits_{n,n',m}e^{it/\hbar \left(E_{n'}-E_{n}\right)} \delta\left(E_0-E_m-\hbar\omega \right) \nonumber\\
&\times \brakete{\Psi_0}{\Psi_{n'}}\bra{\Psi_{n'}}\hat{c}^\dagger_j \ket{\Psi_m}\bra{\Psi_m}\hat{c}_i\ket{\Psi_n}\brakete{\Psi_n}{\Psi_0} \nonumber\\
&= \sum\limits_{m}\bra{\Psi_0(T)}\hat{c}^\dagger_j \ket{\Psi_m}\bra{\Psi_m}\hat{c}_i\ket{\Psi_0(T)}\nonumber\\
&\times \delta\left(E_0-E_m-\hbar\omega \right) \nonumber\\
A^{<}_{ii}(t,\omega)&= \sum\limits_{m}\abs{\bra{\Psi_0(T)}\hat{c}^\dagger_i \ket{\Psi_m}}^2 \delta\left(E_0-E_m-\hbar\omega \right) \nonumber.
}

\bibliography{publication_spectral_function} 

\end{document}